\documentstyle[amsfonts,twocolumn,epsf,aps]{revtex}
\newcommand{\ee}{\end{equation}}
\newcommand{\be}{\begin{equation}}

\begin{document}

\title{Adaptive elastic properties of chromatin fiber}

\author{Eli Ben-Ha\"{\i}m, Annick Lesne and Jean-Marc Victor\footnote{
Email adresses for correspondence: lesne@lptl.jussieu.fr, victor@lptl.jussieu.fr}} 
  
\address{
Laboratoire de Physique Th\'eorique
des Liquides, \\
Universit\'e Pierre et Marie Curie,  \\
Case courrier 121, 4 Place Jussieu, 75252
 Paris Cedex 05,
France\\
}


\maketitle

\abstract{
\small\bf
Chromatin is a complex  of DNA and specific proteins forming
an intermediary level of organization of eukaryotic genomes,
between double-stranded DNA and chromosome.
Within a generic modeling of the chromatin assembly,
we investigate the interplay between the 
 mechanical properties of the  chromatin fiber and its biological
functions. A quantitative step is to relate the mechanics at the DNA level
and the mechanics described at the chromatin fiber level.
It allows to calculate
 the complete set of chromatin  elastic constants
(twist and bend persistence lengths, stretch modulus and twist-stretch coupling
constant), in terms of DNA elastic
 properties and geometric features of the fiber.
These elastic  constants are strongly  sensitive
to the local architecture of the fiber and we
 argue that this tunable elasticity might be a key feature in chromatin
 functions, for instance in the initiation and regulation of transcription.
Moreover, this analysis provides a framework to interpret micromanipulations
studies of chromatin fiber and suggests further experiments involving intercalators
to scan the tunable elasticity of the fiber. 
}



\vskip 10mm
\noindent{\bf Introduction}
\vskip 2mm
Chromatin is a  protein-DNA complex
forming the basic material of  chromosomes of all 
eukaryotic organisms 
and strikingly conserved during evolution \cite{Wolffe}.
It provides an intermediary level of organization
between the underlying DNA double-helix and the whole cell
machinery.
Chromatin  achieves essentially two fundamental functions:
it  compacts efficiently DNA into the cell nucleus, and it
 compacts it operationally in order to allow targeted
 gene expression.
We propose a mechanical approach of these two questions.
It 
suggests that the chromatin structure
might be selected according to its elastic properties,
which  might be exploited during transcription by the various enzymes at work during 
this process, for instance to locally condense (gene silencing) 
or decondense (gene activation) the fiber.

\vskip 6mm
\noindent{\bf Chromatin  assembly}
\vskip 2mm


We  introduce a basic geometric modeling of chromatin assembly, 
which allows 
 to incorporate acknowledged data on  DNA structure, its bend 
and twist persistence lengths, and data on the nucleosome structure
 into an analytically tractable framework.
Chromatin is composed of a double-stranded DNA molecule wrapped from place to
place around protein cores (histone octamers).
Focusing on the native case where these cores 
are fixed on the DNA, we model separately the nucleosomes, i.e.
the DNA-wrapped histone cores, and the linkers, 
 i.e. the naked DNA segments connecting the nucleosomes.

Since we are looking for generic properties of the chromatin fiber, originating from
its assembly, sequence effects are ignored.
We  suppose   that all the linkers have the same
number  $n$  of base pairs;
it corresponds to currently observed phased nucleosomes
\cite{yao}.
{\it A posteriori}, we may argue that self-organization
of the fiber  architecture leads to such a regular
positioning of the nucleosomes in order to get
proper structural and elastic properties.
In fact, we need this hypothesis to be satisfied only  locally, over a
few linkers.
Linker DNA is 
described  as a non extensible semi-flexible polymer
 within the continuous 
worm-like-chain model supplemented with twist elastic degree
 of freedom \cite{Bouchiat}; it thus involves 
two persistence lengths, respectively  of bend ($A=53$ nm) and twist
($C=75$ nm), the given  values corresponding to 10mM NaCl \cite{croquette}. 
Linkers are straight in absence of applied
constraints.

Nucleosome
structure is now well-known thanks to high resolution
cristallography data\cite{Luger}.
The net
effect of a nucleosome on the DNA trail can be described as a rigid 
 kink connecting two successive linkers.
The slope of the DNA path on a nucleosome being fixed 
(angle $4.5^o$), a single 
 angle $\Phi$ between entering and outgoing linkers
thoroughly prescribes the geometry.
This picture allows  an effective modeling of various physico-chemical effects,
for instance
electrostatic repulsion between linkers and its variable
screening when the ionic strength is increased. It accounts also for
histone H1 binding in the neighborhood
of linker entry/exit points as well as histone tails influence and its variation
upon posttraductional covalent modifications of these tails
\cite{leuba-H1}\cite{bednar-H1}\cite{H1-thomas}\cite{H1-holde}.
We here ignore interactions between nucleosomes, which  is valid 
as soon as the internucleosomal
distances are larger than the interaction range;
in any cases, the model describes the geometric contribution
to chromatin elastic behavior
originating from the DNA elastic
properties.

Our model,
quite similar to the so-called two-angle model
of Woodcock et al. \cite{struct-4}, amounts to build  
the chromatin
fiber nucleosome after nucleosome.
The rotational positioning of a  nucleosome  with respect  to
the previous one
 is entirely prescribed by the twist angle of the DNA linker
connecting them.
In the relaxed state, it is equivalently prescribed  by
the linker length $n$ (in bp) since the twist angle is then precisely 
equal to $\tau=2\pi\; n/10.6$ (recall that there is 10.6 bp per  DNA turn).
Each step of the assembly is described analytically, but  
we implemented this model within a Maple program in order to handle an arbitrary number
of nucleosomes and to perform a quantitative analysis of the 
chromatin fiber geometric properties.


\vspace*{10mm}

\leavevmode
\epsfxsize= 15pt
\epsffile[100 150 200  700]{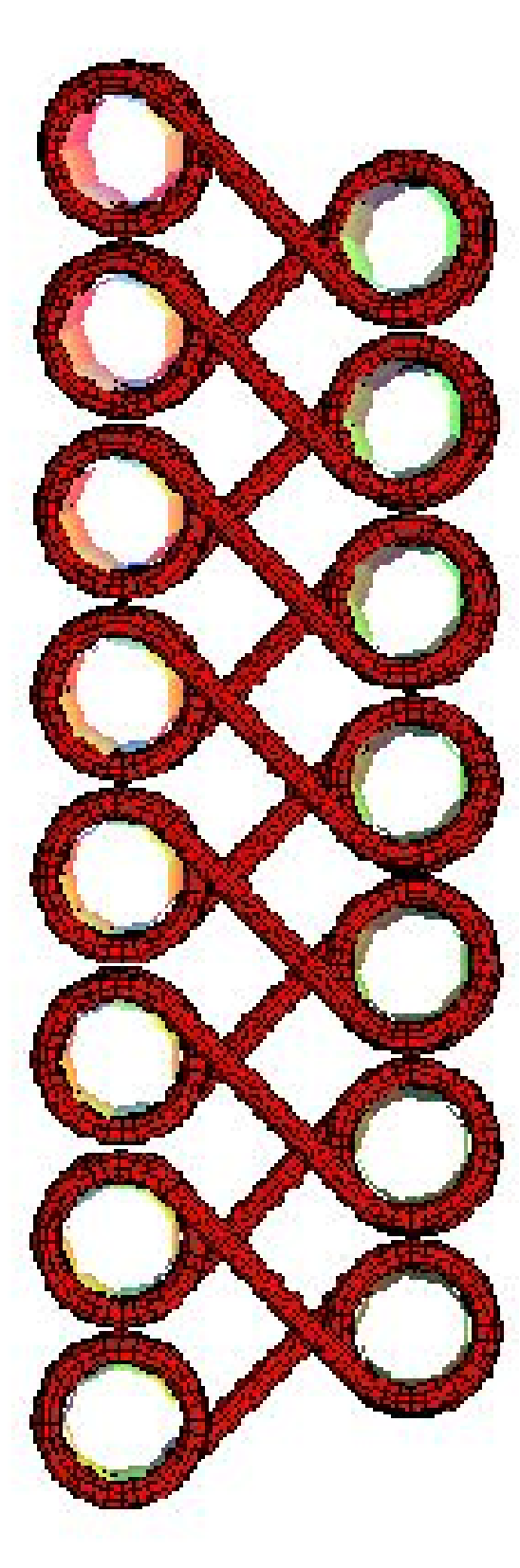}
\hspace{30mm}
\epsfxsize= 15pt
\epsffile[100 150 200  700]{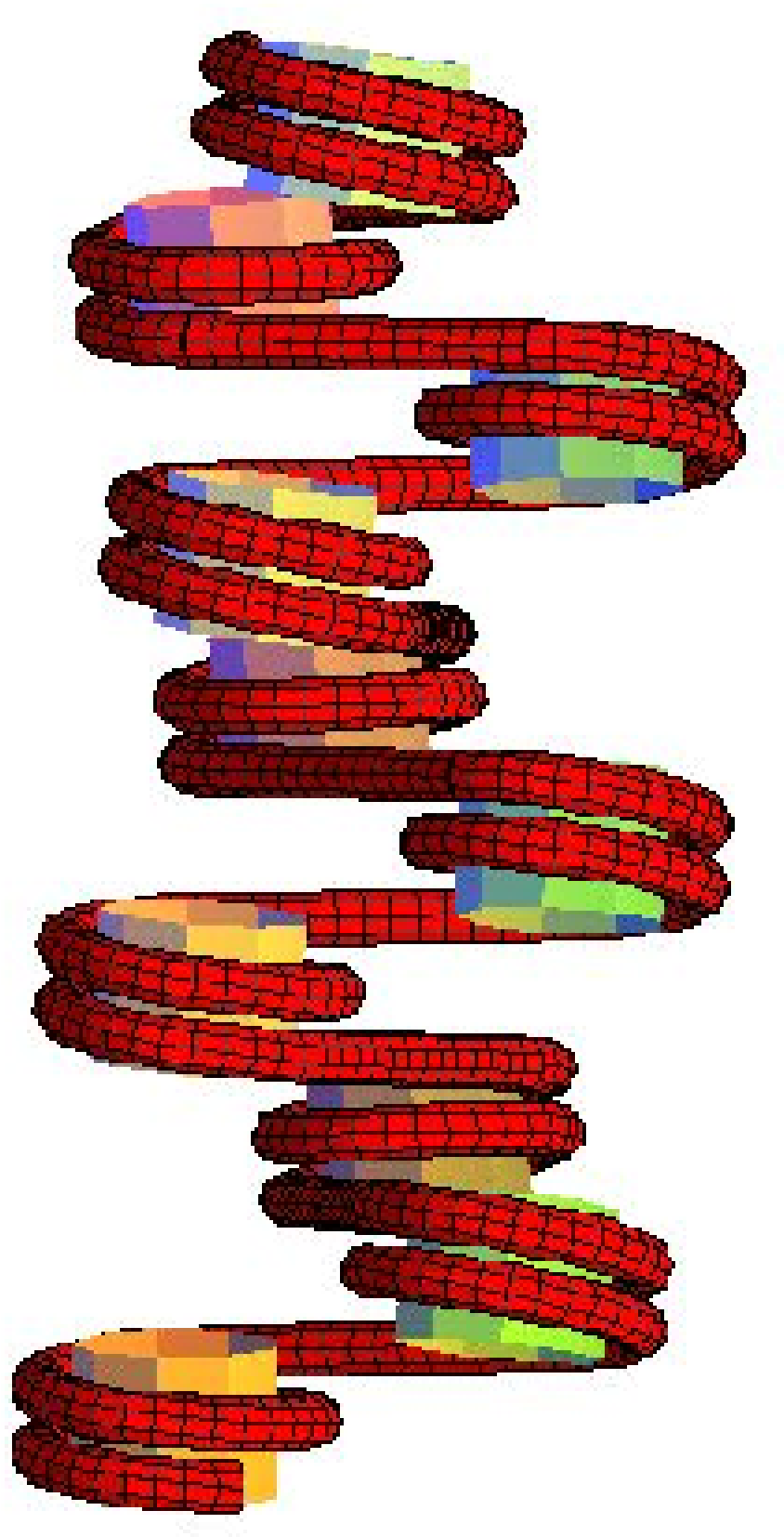}

\vspace*{15mm}
\leavevmode
\hspace*{15mm}
\epsfxsize= 10pt
\epsffile[100 150 200  700]{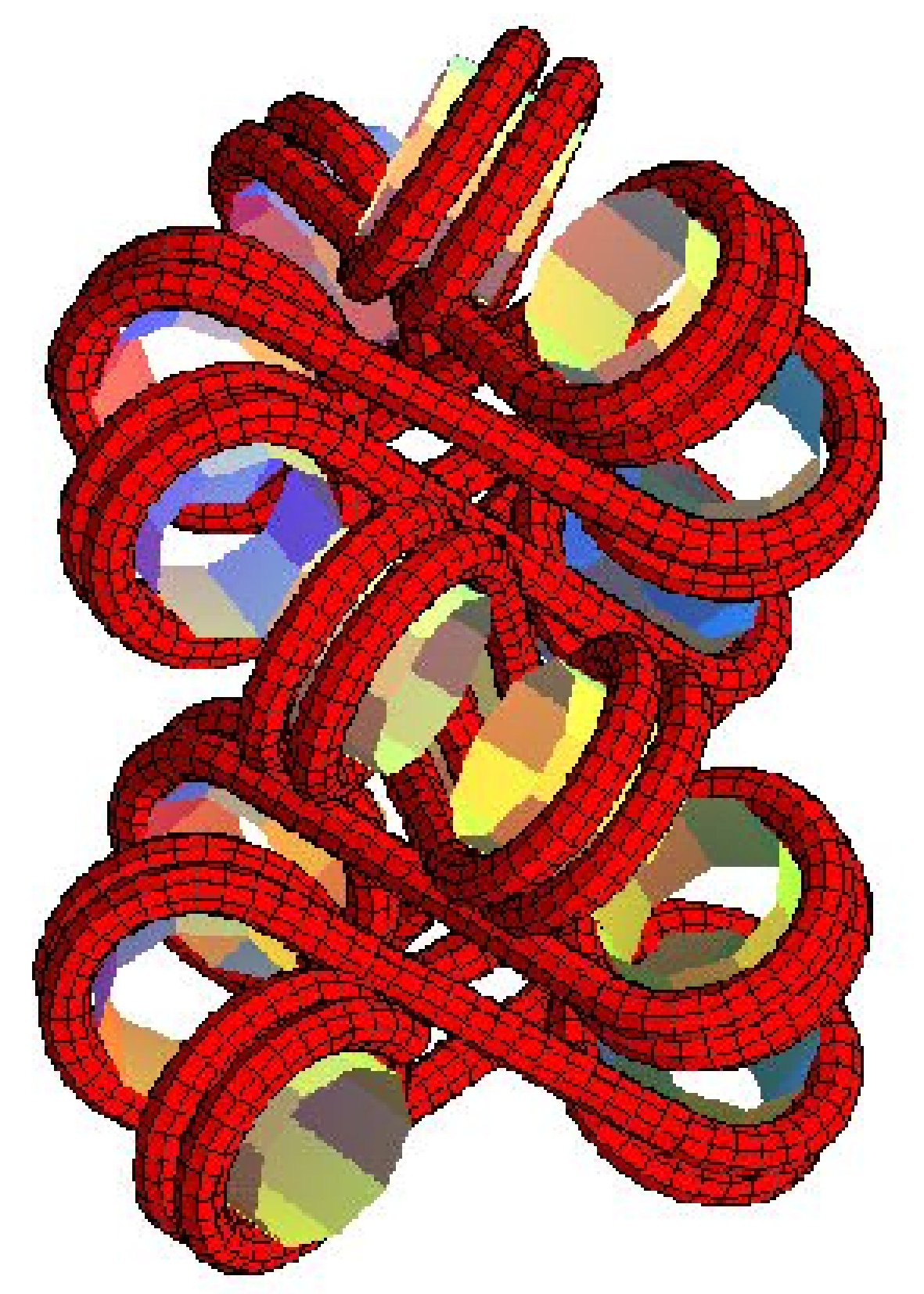}

\vskip 8mm
\noindent{\bf  Figure 1:}
Varying the linker length $n$ and
the entry/exit angle  $\Phi$ recovers all the
structural  diversity observed experimentally, for instance
 ribbon ($\Phi\!=\!90^o$, $n\!=\!58$ bp),
 columnar structure ($\Phi\!=\!50^o$, $n\!=\!42$ bp)
or  double-helical structure ($\Phi\!=\!60^o$, $n\!=\!56$ bp).

\vskip 6mm
\noindent{\bf Structural properties}
\vskip 2mm

The implementation of the above model
straightforwardly 
  gives all the geometric properties of the chromatin fiber.
Homogeneity of the local architecture (uniform
values for $n$ and $\Phi$) enforces
a helical symmetric structure for the fiber; we shall speak of
``chromatin superhelix'' (SH), and denote $\vec{A}$ its axis.
A complete study of the symmetry properties and 
  quantities
relevant to the mechanics of the fiber has been presented in a previous
paper \cite{BLV}.
We here only underline the high structural sensitivity
of the fiber, evidenced on Figure 1:
for various values of $n$ and $\Phi$,
we obtain the whole variety of structures suggested from  experimental studies
\cite{Wolffe} \cite{struct-1} \cite{struct-2} \cite{revuewidom}. These qualitative results support our minimal 
two-parameter modeling, since it is sufficient
to recover  the 
 actual diversity of chromatin structures.
A noticeable point, evidenced on Figure 2,  is the fact that the diameter
remains about 30 nm whatever the structure; the so-called 30nm-fiber
observed experimentally might thus correspond to far different structures.
Excluded volume has to be taken into account
to discard unrealistic structures; 
we determined a sufficient condition to
avoid steric hindrance,  which is shown on Figure 2.

One of the conclusions of our  structural study is the fact 
that the connection  between the microscopic parameters and the SH geometric
characteristics  is too complex and multivariate to get 
small-scale informations from structural
observations of the fiber; we now  turn to the analysis
of its mechanical properties.

\vspace*{-60mm}

\leavevmode

\hspace*{10mm}
\epsfxsize= 25pt
\epsffile[100 150 200  700]{fig2.eps}

\vskip 70mm
\noindent{\bf  Figure 2:} Diameter of the chromatin fiber ({\it upper blue  curve})
and its pitch ${\cal P}$ ({\it lower red  curve}).
The thin green  line gives a lower bound ${\cal P}_c$
for the pitch above which no steric hidrance occurs within the fiber,
in constrast to the dashed regions where ${\cal P}<{\cal P}_c$.

\vskip 6mm
\noindent{\bf Elastic properties}
\vskip 2mm

In order to obtain the leading order of the elastic constants
of the fiber, we merely consider a straight fiber with cylindrical symmetry.
Describing the elastic properties of
the chromatin fiber is thus nothing but a problem of spring mechanics \cite{love}.
Nevertheless, the architecture of this ``spring'' is much more complex than a simple
helical coiling and we expect that the detailed structural
 features of the chromatin 
assembly  strongly influence the  behavior at the fiber scale.
Our modeling allows us to  compute analytically
 the elastic constants describing  the
linear response of the fiber
when a force $\vec{F}$ and a torque $\vec{M}$
are applied at its ends,  hence to 
investigate quantitatively how the elastic response 
of the fiber  varies with its relaxed  structure. Note that this 
structure is  itself
controlled by the ``microscopic'' structural parameters $n$ and $\Phi$.

Our computation  lays on the  identification
of  the elastic energy
of the fiber when a force $\vec{F}$ and a torque $\vec{M}$
are applied at its ends 
 and the sum of the elastic energies  of its linkers (bend and twist energies)
in such a constrained configuration.
This identity follows from the
absence of internucleosomal interactions
and to our focus on the linear response regime
 (no modification
of the fiber architecture at this order);
it amounts to restrict to the elastic contribution
originating from linker elasticity within the relaxed chromatin structure.
The first step  is then  to  
  relate  the ``macroscopic'' stresses $(\vec{F},\vec{M})$
to the distribution of stresses  along linker DNA.
We denote $\vec{f}_j(s)$ the force
and  $\vec{m}_j(s)$ the torque  exerted at the point $P_j(s)$ of linker $j$
(with arclength $s$)
by the upstream part of the fiber.
We focus on the universal, rotationally symmetric behaviour of the fiber.
Writing  standard
 equilibrium equations of spring mechanics,
and using the linear response hypothesis
to simply sum up the effects of the  different stresses
applied to
the fiber
(stretching force, bending torque and torsional torque), we obtain:
\vskip 4mm
\be\label{fF}
\mbox{
\begin{tabular}{|rcl|}
\hline
&&\\
\hspace{10mm}$\vec{f}_j(s)$&=&$\vec{F}$\\&&\\
 $\vec{m}_j(s)$&=&$\vec{M}-[\vec{O_j(s) P_j(s) }]\wedge\vec{F}$\hspace{10mm}
\\&&\\
\hline
\end{tabular}}\ee
\vskip 2mm
\noindent
where $O_j(s)$ denotes  the orthogonal projection of $P_j(s)$ onto the SH axis 
$\vec{A}$.
It gives the relation between the global stresses $\vec{F}$ and $\vec{M}$
exerted at the fiber ends and the local stresses 
$\vec{f}_j(s)$ and $\vec{m}_j(s)$ experienced at the linker level, at
each point of the DNA path.
The term $[\vec{O_j(s) P_j(s) }]\wedge\vec{F}$ reflects the involvement of the fiber
architecture in the expression of the local torque $\vec{m}_j(s)$.
This set of equations is the core of our analysis since it 
allows to perform explicitly the change of  description level,
namely passing from the DNA level to the fiber level and reciprocally.

As we restrict to the description of harmonic elasticity, the coefficients of 
$\vec{F}$ and $\vec{M}$ in the elastic energies of the linker will be computed within the
relaxed SH.
We underline that we do not need to compute
the constrained shape of the linkers to describe the linear response
of the fiber to applied force and torque.
The cylindrical symmetry of the fiber 
ensures that $\vec{F}$ is directed along the SH axis $\vec{A}$ at
equilibrium.
 $\vec{M}$ decomposes into a bending torque 
$\vec{M}_b$ and a torsional torque $M_t$ according to 
$\vec{M}=\vec{M}_b+M_t\vec{A}$.
The well-established 
  elastic energy of a linker \cite{Bouchiat}\cite{croquette} can then be 
expressed 
as a  function of $F$, $M_t$ and $M_b$ thanks to (\ref{fF});
it yields a quadratic expression
\be
{\cal E}_{\rm linker}=
a_sF^2+a_tM_t^2+2a_{ts}M_tF+a_bM_b^2
\ee
 which is to 
be identified with the elastic energy 
${\cal E}_{\rm fiber}(F,M_t, M_b)$ of a length $D$
of chromatin fiber, where $D$ is the distance between two successive
nucleosomes measured along the SH axis (projected distance).
Plugging in the quadratic expression ${\cal E}_{\rm linker}(F,M_t, M_b)$
the  linear  response ansatz relating 
 the  stresses $F$, $M_t$ and $M_b$  experienced by the fiber
and its  strains $u$
(relative extension), $\Omega$ (twist rate) and $\rho$ 
(local curvature):
\be\label{linearansatz}
\left(\begin{array}{l}
F\\M_t\\M_b
\end{array}\right)=
\left(
\begin{array}{ccc}
\gamma&k_BTg&0\\
k_BTg&k_BT{\cal C}&0\\
0&0&k_BT{\cal A}
\end{array}
\right)
\left(
\begin{array}{l}
u\\ \Omega\\\rho
\end{array}\right)
\ee
naturally embeds  the fiber and its
elastic behaviour within  a current continuous model
known as the ``extensible worm-like rope'' ( EWLR)\cite{marko-stretch-2},
in which the density of elastic energy writes,
as a function of the canonical variables $u$, $\Omega$ and $\rho$:
\be\label{EWLR1}
\epsilon_{EWLR}={k_BT{\cal A}\varrho^2\over 2}+{k_BT{\cal C}\Omega^2\over 2}+
{\gamma\,u^2\over 2}
 +k_BTg\,\Omega u
\ee 
${\cal A}$ is the bend persistence length of the fiber, ${\cal C}$
its twist persistence length, $\gamma$ its stretch modulus
(dimension of a force) and $g$ the twist-stretch
coupling constant (no dimension).
Note  the expected vanishing of the twist-bend and stretch-bend
coupling (for symmetry reasons\cite{marko-stretch-2}. 
We may now express the elastic constants of the fiber
 in terms of the coefficients $a_b$,  $a_t$,
$a_{ts}$  and $a_s$, obtained as functions of the DNA elastic constants
and fiber geometry, by a term-wise identification of the coefficients
of $u^2$, $\Omega^2$, $u\Omega$ and $\rho^2$ in the following identity:
$$
{\cal E}_{\rm fiber}\equiv D\;\left({k_BT{\cal A}\varrho^2\over 2}+{k_BT{\cal C}\Omega^2\over 2}+
{\gamma\,u^2\over 2}
 +k_BTg\,\Omega u
\right)
$$
\be={\cal E}_{\rm linker}[F(u,\Omega),M_t(u,\Omega), M_b(\rho)]
\ee
from which follows analytical expression\cite{BLV} of the elastic constants
${\cal A}$, ${\cal C}$, $g$ and $\gamma$.
We  present
on Figure 3  the results for $\Phi=50^o$ (high salt concentration,
 with histone H1,
i.e.  physiological conditions) when the linker length is varied.
The curves  evidence a strong sensitivity of elastic coefficients
 with respect to the fiber structure, here controlled by $n$.
This sensitivity  originates in the sensitivity of the SH pitch:
accumulation of turns in a spring leads to a
highly flexible  accordion-like behavior.
It is to note that the persistence length ${\cal A}$ and ${\cal C}$ are almost identical,
 both close to the SH pitch and smaller than the corresponding
persistence lengths $A$ and $C$ of DNA.
More strinkingly, the stretch modulus of the chromatin fiber,
around 5 pN, is to be compared to that of DNA ($\gamma_{\rm DNA}\approx 1100$ pN):
whereas DNA can be seen as non extensible, chromatin is highly
extensible; in particular, entropic and enthalpic elastic regimes (well-separated 
in the case of DNA) will overlap.
We finally underline a key feature:
the twist-stretch coupling constant, reflecting the chirality 
of the chromatin fiber, exhibits steep variations and sign changes when the 
linker length is varied.
The same twisting of  the fiber
will either condense or decondense it according to the relaxed structure.


\vskip 7mm
\noindent{\bf Experimental implications}
\vskip 2mm

A first issue involving 
fiber elasticity 
 is to describe the linear elastic response of the fiber to global
stresses, i.e. a force and a torque applied at its ends.
This issue refers to micromanipulations in which a single chromatin fiber
is pulled \cite{cui-bust}  (and possibly will be twisted).
Describing the fiber within the EWLR model, yet thoroughly
investigated in the context of DNA by Marko\cite{marko-stretch-2}, allows 
an interpretation of the force-extension curves in terms of geometric and
mechanical parameters of the  fiber.
Experimental results of Cui and Bustamante \cite{cui-bust}
are in good agreement with our predictions
(note that their measure rests on relaxation curves, for which there 
is no interaction between  nucleosomes, consistently with our modeling).
Indeed, they measured a bend persistence length 
${\cal A}\approx$ 30 nm and a stretch modulus $\gamma\approx 5$ pN,
which are obtained simultaneously in our modeling for 
$n\approx 42$, or any integer multiple of DNA pitch;
we there recover the nucleosome phasing observed in biological experiments
\cite{yao}.

 Our study should now be supplemented by 
investigating the effect of a dispersion in the values of $n$ and 
$\Phi$.
It is likely that the dispersion will
induce  a kind of averaging.
Continuity arguments lead us to expect the sensitivity of the
 structural and elastic properties to remain -- although possibly smoothed out --
together with its possible relevance
 in biological regulation.
Work is in progress on this question.

To circumvent this limitation, we suggest to take advantage
 of the interplay between linker intercalation and the
chromatin fiber mechanics.
The twist change $\Delta\tau$
achieved in the linker by intercalator molecules
(for instance with ethidium bromide {\it in vitro})
amounts to shift the linker length  by
$\Delta n=10.6 \times \Delta\tau/2\pi$.
Plotting the various elastic constants as a function of $\Delta n$ 
should provide similar curves as those shown on Figure 3, however smoothed out.
We therefore suggest that it 
 should be possible to scan experimentally these curves
by varying continuously the concentration of intercalators
added in the buffer when performing pulling fiber experiments.

\noindent

\vspace*{-55mm}

\leavevmode
\hspace*{10mm}
\epsfxsize= 25pt
\epsffile[100 150 200  700]{fig3a.eps}

\vspace{5mm}
\leavevmode
\hspace*{10mm}
\epsfxsize= 25pt
\epsffile[100 150 200  700]{fig3b.eps}

\vskip 70mm
\noindent{\bf  Figure 3:} Elastic constants of the
chromatin fiber versus linker length $n$ (in bp) for $\Phi=50^o$.
A similar sensitivity 
when $n$ varies is observed for other values of $\Phi$,
or when varying $\Phi$ at fixed $n$.

\noindent
{\bf (a)} \ 
Bend persistence length
${\cal A}$ ({\it solid red line}) in nm and stretch modulus in pN
({\it dashed blue line}); here is shown the effective
 modulus  $\gamma_{eff}=\gamma-k_BTg^2/{\cal C}$ which is involved 
in  force-extension curves\cite{marko-stretch-2}\cite{BLV} hence directly measurable.
Our computations predict  that ${\cal A}$ and the twist 
persistence length ${\cal C}$ are almost equal, both close to the
pitch ${\cal P}$ of the SH (Figure 2). 
Experimental measurements of Cui and Bustamante\cite{cui-bust}
yield a value about 30 nm for  ${\cal A}$  and
 5 pN for  $\gamma_{eff}$ ({\it horizontal lines}). 
 
\noindent
{\bf (b)} \  Twist-stretch coupling constant $g$ (no dimension).
Note that the vanishing of 
$g$ corresponds to  a change of chirality for the fiber.


\vskip 7mm
\noindent{\bf Biological relevance}
\vskip 2mm
Another   issue, directly relevant to the {\it in vivo}
functioning of chromatin, is  to investigate both the effect  of
DNA-protein binding on the  fiber structure, and the
effect of fiber stresses on the linker DNA-protein binding.
Our theoretical approach  gives a framework to such studies 
by relating the structure and mechanics
at the DNA scale and those at  fiber scale.

On one hand, our approach   allows  to describe the response 
of the fiber to local, internal stresses as
those created by intercalators, groove-binding proteins, or
 any induced change in the
fiber assembly  at the linker scale.
Mechanical sensitivity is likely to provide efficient switches for processes
involving a conformational transition  of the fiber, for 
instance the decondensation required prior transcription or, at the opposite,
fiber compaction currently associated with gene silencing \cite{widom2}.
A possible tuning mechanism lays on the variation of $\Phi$, controlled in particular
by the
presence of linker histone,  salt concentration and histone tails binding affinities.

On the other hand,  protein-DNA interactions involved 
in gene-regulated subcellular events take place
within   chromatin fiber.
In consequence, they are likely to be influenced, if not directly
regulated, by  chromatin structure
and by the stresses generated in the linkers when the chromatin 
fiber is somehow constrained.
In particular, we expect such an involvement of chromatin structure 
and multilevel
mechanical properties to be at work in epigenetics and gene
imprinting \cite{science}.

\vskip 6mm
\noindent{\bf Conclusion}
\vskip 2mm
The chromatin scale is precisely  the scale of nanomechanics:
at this scale, we expect a strong and direct interplay between the biological
functioning, monitored  by various enzymes, and the mechanical
properties of the chromatin fiber.
Our complete and quantitative analysis gives clues 
to discuss how the elastic properties of the chromatin fiber
might at the same time
 favour DNA compaction into the
chromosomes and 
 allow 
local  and controlled decondensation of chromatin involved in gene expression.


\end{document}